# Reengineering PDF-Based Documents Targeting Complex Software Specifications[*]

MEHRDAD NOJOUMIAN[1]
University of Waterloo, Canada
and
TIMOTHY C. LETHBRIDGE[2]
University of Ottawa, Canada

This article aims at reengineering of PDF-based complex documents, where specifications of the Object Management Group (OMG) are our initial targets. Our motivation is that such specifications are dense and intricate to use, and tend to have complicated structures. Our objective is therefore to create an approach that allows us to reengineer PDF-based documents, and to illustrate how to make more usable versions of electronic documents (such as specifications, technical books, etc) so that end users to have a better experience with them. The first step was to extract the logical structure of the document in a meaningful XML format for subsequent processing. Our initial assumption was that, many key concepts of a document are expressed in this structure. In the next phase, we created a multilayer hypertext version of the document to facilitate browsing and navigating. Although we initially focused on OMG software specifications, we chose a general approach for different phases of our work including format conversions, logical structure extraction, text extraction, multilayer hypertext generation, and concept exploration. As a consequence, we can process other complex documents to achieve our goals.

## 1. INTRODUCTION

Published electronic documents, such as specifications, are rich in knowledge, but that knowledge is often complex and only partially structured. As a result, it is usually difficult for users to make maximum use of a document. The objective of this research is to develop an approach by which a typical published specification can be made more usable to end-users. We achieve this by reengineering the PDF version of a document in order to generate a new multilayer hypertext version of that document. This makes the knowledge more explicit, and facilitates searching, browsing, navigating, and other operations required by end users.

As a case study, we applied our approach to various OMG software specifications published in PDF format. However, we ensured that all aspects of our work are as general as possible so that the same approach can be applied to other documents. We chose OMG specifications because they (a) have particularly complicated structures, (b) are important to the software engineering community, and (c) have been studied in depth by members of our research group who have experienced frustration with them.

Our overall approach is an example of document engineering, and is divided into two distinct phases. The first step is to extract the document's logical structure and core knowledge, i.e., representing the document in a meaningful XML format [Nojoumian and Lethbridge 2007]. This result consists of content information and excludes irrelevant details of the original document's presentation. Capturing the content in XML allows for easy exploration and editing of data by XML editors and other tools, and allows the generation of the new presentation to be a separate responsibility. The initial task in phase one is to use a Commercial Off-The-Shelf (COTS) tool to convert an input PDF file into a format we can more readily work with. We conducted an experiment to see which tool would generate the best XML version of the document.

---

[1] David R. Cheriton School of Computer Science, University of Waterloo, Waterloo, Canada, mnojoumi@cs.uwaterloo.ca

[2] School of Electrical Engineering and Computer Science, University of Ottawa, Ottawa, Canada, tcl@site.uottawa.ca

The next task was to parse the output of the COTS tool to clean up the XML file and create meaningful tags based on the document headings, i.e., Table of Contents (ToC), for further processing. The second phase of our approach is to construct a multilayer hypertext version of the document in order to make a complex document more usable by allowing navigation of both its structure, and also of semantics described by the document. We believe that if developers of specifications publish their documents in the format that we developed, it will greatly assist end users of software specifications.

## 1.1 Motivation and Contribution

The motivation for our work is that complex documents such as software specifications are not as usable as we believe they should be. By *complex* we refer to a document which has most of the following features: a large number of pages and figures, interconnected concepts, definitions or equations spread throughout the document, numerous cross references, intricate tables expanded over successive pages with figures and hyperlinks in their cells, nested lists with complicated hierarchical structures, and long samples of programming code spread over page boundaries.

In other words, they are large, dense and intricate to use, so most users will skim them or look things up when needed. However, readers often have to jump backwards and forwards many times to follow cross-references. Numerous concepts tend to be connected only implicitly; it is not easy for end users to follow references to the place where the reference points. For instance, in the UML specifications, there are definitions of metaclasses. Each of these has inherited properties coming from metaclasses that may be in other packages.

Nowadays, documents are published using a format that mimics legacy paper documents. Although PDF is an excellent way of rendering a paper document faithfully in electronic form and has some built-in navigation capability, the use of PDF takes away some potential usability and makes the access to its structured content difficult; requiring reverse engineering techniques. For this reason, various pre-processing tools have been developed to allow the extraction of a PDF file's textual content. However, these tools have limited capabilities in the sense that the text's reading order is not necessary preserved, especially when dealing with complex layouts [Hadjar et al. 2004].

The above issues raise the following research questions: How can we reengineer a PDF-based document in as general and straightforward way as possible? What facilities are required for end users to have a better experience with a document? The following is a list and brief description of the key contributions of our work for document engineering.

**An efficient technique for capturing document structure:** we experimented with conversions using different COTS tools to select the best file transformation, i.e., extracted document's logical structure in a clean XML format. We further processed this using a parser written in Java. We encountered problems such as mis-tagging related to the conversion phase and lack of well-formed characteristic of our XML file. We overcame these problems and generated a well-formed XML document with various types of meaningful tags, which facilitated our further processing.

**Various techniques for text extraction:** we experimented with numerous methods to create a usable multilayer hypertext version of the document for end users. We also applied the latest W3C (World Wide Web Consortium) technologies for concept extraction and cross-referencing to improve the usability of the final output.

**A general approach for document engineering:** although our targeted documents were OMG specifications, we chose a generic approach for various phases of our work including format conversions, logical structure extraction, text extraction, hypertext generation, and concept exploration. As a result, we can process other complex documents. We also established the major infrastructure of a document-engineering tool.

**Significant values and usability in the final result:** after showing how to create a more useful format of a document, we demonstrate the usability of our final outcome such as: better navigating and scrolling structure, simple textual content processing, efficient learning, faster downloading, as well as easier printing, monitoring, coloring, and cross referencing.

## 2. RELATED WORK

In this section, we review document structure analysis and some research with respect to analyzing PDF documents and leveraging tables of contents.

### 2.1 Document Structure Analysis

Klink et al. [2000] present a hybrid and comprehensive approach to document structure analysis. Their approach is hybrid in the sense that it makes use of layout (geometrical) as well as textual features (logical) of a given document. Mao et al. [2003] propose numerous algorithms to analyze the physical layout and logical structure of document images (images of paper documents) in many different domains. The authors provide a detailed survey of diverse algorithms in the following three aspects: physical layout representation, logical structure representation, and performance evaluation.

Summers [1998] explains an approach for finding a logical hierarchy in a generic text document based on layout information. The logical structure detection has two stages, segmentation and classification. The first one separates the text into logical pieces, and its algorithm relies totally on layout-based cues, while the second one labels the pieces with structure types, and its algorithm uses word-based information. Tsujimoto and Asada [1990] represent a document's physical layout and logical structure as trees. They characterize document understanding as transformation of a physical tree into a logical one. Blocks in the physical tree are classified into head and body while in the logical tree are categorized into title, abstract, sub-title, paragraph, page number, caption, etc.

Lee et al. [2003] provide a syntactic method for sophisticated logical structure analysis, which transforms multiple-page document images with hierarchical structure into an electronic document in XML. Their proposed parsing method takes text regions with hierarchical structure as input. Conway [1993] uses page grammars and page parsing techniques to recognize document logical structure from physical layout. The physical layout is described by a set of grammar rules. Each of these rules is a string of elements specified by a neighbor relationship such as above, left-of, over, left-side, and close-to. For describing the logical structure a context-free string grammar is used.

Aiello et al. [2000] provide a framework for analyzing colored documents of complex layout. In this framework, no assumption is made about the layout. The proposed structure combines two major sources of information: textual and spatial. It also uses shallow natural language processing tools (such as partial parsers) to analyze the text.

## 2.2 PDF Document Analysis

Anjewierden [2001] describes a method in order to extract the logical structure from PDF documents. The approach is based on the idea that the layout structure contains cues about the logical structure, for instance, a text object in a large bold font. As a result, the logical structure (e.g., a section title or heading) can be detected incrementally.

Chao and Fan [2004] develop various techniques that discover different logical components on a PDF document page. They first partition a page into text blocks, images blocks, vector graphics blocks, and compound blocks. They then present the results of this analysis in an XML format. Hadjar et al. [2004] propose a new approach for the extraction of the document content by low-level extraction techniques applied on PDF files as well as layout analysis performed on TIFF images. They first illustrate various steps of their method, and then present a first experiment on the reconstruction of the newspapers' reading order.

Rigamonti et al. [2005] demonstrate a reverse engineering tool for PDF documents, named "Xed". This tool extracts the original document layout structure in a hierarchical canonical form (i.e., independent of the document type) by means of electronic extraction methods and document analysis techniques. Bloechle et al. [2006] present different approaches for processing the structured content of PDF documents based on image analysis and electronic content extraction. They also demonstrate an algorithm for restructuring a document in XCDF (eXhaustive Canonical Document Format), which is based on XML and has well-defined properties facilitating access to the structured content.

## 2.3 Leveraging Tables of Contents

Dejean and Meunier [2005] describe a technique for structuring documents according to the information in their tables of contents (ToC). In fact, the detection of a ToC, as well as the determination of the parts it refers to in the document body, rely on a series of properties that characterize any ToC. He et al. [2004] propose a new technique for extracting the logical structure of documents by combining spatial and semantic information of the table of contents. They exploit page numbers and numbering schemes to compute the logical structure of a book. Their method is not a general approach because of the observed diversity of page or section numbering, and of ToC layout.

Lin et al. [1997] propose a method for analyzing the logical structure of books based on their tables of contents by layout modeling and headline matching. In general, the contents page holds accurate logical structure descriptions of the whole book. In this approach, text lines are first extracted from the contents page, and OCR (Optical Character Recognition) is then executed for each text line. The structures of the page number, head, foot, headline, chart, and main text of the text page are analyzed and matched with information obtained from the contents page.

Belaid [2001] presents a labeling approach for the automatic recognition of a table of contents. A prototype, called "Calliope", is applied for electronic consulting of scientific papers. This approach works on structured ASCII files produced by OCR. Lin and Xiong [2005] introduce a new approach to explore and analyze a ToC based on content association. Their method leverages the text information in the whole document, and it can be applied to a wide variety of documents without analyzing the model of an individual document; NLP (Natural Language Processing) and layout analysis are integrated to improve the ToC tagging. Bourgeois et al. [2001] describe a statistical model for document understanding, which uses both text attributes and document layout. In this model, probabilistic relaxation (which is a general method to classify objects and to repetitively adjust the classification) is used as a recognition method for understanding the table of contents and discovering the logical structure.

## 3. DOCUMENT TRANSFORMATION

We selected PDF-based documents for processing for the following reasons. First of all, people do not have access to the original word-processor formats of many documents. When a document is published to the web, an explicit choice is usually made to render the result as PDF or HTML to guarantee that everyone can read it without having to have Microsoft Word, FrameMaker, etc. In addition, the PDF format has useful features that make it semi-structured. For instance, it often contains bookmarks created from headings to enable a user to navigate a document; a computer can also use this information to extract the logical structure. Finally, since a PDF file can be easily generated from most document formats, there exist a huge number of PDF documents on the Internet.

One of our major goals is to extract the document's logical structure. As we mentioned earlier, many key concepts of the targeted OMG specifications are expressed in the logical structure. By extracting this structure and representing it as XML, we can form an excellent infrastructure for our subsequent processing. We solved this problem in two phases. In this section, we describe the first step, i.e., transforming the raw input into a format more amenable to analysis. The second step, i.e., extracting and refining the logical structure, is the topic of the next section.

To extract the logical structure of a document, we performed various conversion experiments using different tools such as *Adobe Acrobat Professional*, *Microsoft Word*, *Stylus Studio® XML Enterprise Suite*, and *ABBYY PDF Transformer* to see to what extent each could facilitate the extraction process.

### 3.1 Criteria

Since we want to extract the document's logical structure and convert it to XML, we are interested in an output format that can most facilitate this extraction. To select the best conversion, we defined a set of criteria based on the experiences we gained during our experiments. These criteria are as follows:

- **Generality**: A format should enable the design of a general extraction algorithm for processing other electronic documents.
- **Low volume**: We should avoid a format consisting of various extra materials not related to the document content. This includes information related to the presentation format, for instance, the position of elements such as words, lists and paragraphs.
- **Easy processing**: Even if a format results in a small file, it still may not be adequate. It should also be clean and machine-readable. For instance, formats that purely mark constructs such as paragraphs with a single marker are easier to work with than formats that do not do this.

- **Tagging structure**: We prefer a format that has a tagging structure, such as XML or HTML, because we want our final output of this step to be a structured format.
- **Containing clues**: A format should use markers, which provide accurate and useful clues for processing and finding the logical structure. For example, meaningful keywords regarding the headings: "LinkTarget", "DIV", "Sect", "Part", etc.

Sometimes, formats that contain extra data such as font, size, style, and position are more helpful, while in other cases documents that are mostly text without additional details would be more useful. For instance, the extra data would be useful for algorithms that detect headings of a document based on this information, whereas style and font tags are of little use to our algorithm. Hence, we would like to compromise among different kinds of formats to satisfy our mentioned criteria. In the next part, we evaluate different transformations to define the best candidate.

### 3.2 Evaluation

To narrow down the list of possible transformations, we evaluated each transformation according to how it satisfies our criteria. We performed all conversions on various OMG software specifications with different tools. Our observations are as follows:

DOC and RTF formats are generally messy. For example, they code figures among the contents of the document, whereas some formats such as HTML or XML put all the figures in a separate folder in an image format. In addition, DOC and RTF store information related to the font, size, and style of each heading, paragraph, and sentence beside them. This information is not useful for us because it varies from one document to another, contradicting the generality property and increasing the potential for noise during processing. TXT format is very simple but does not give us consistent clues about where to find the beginning of chapters, headings, tables, etc.

PDF is complex, but after Adobe Acrobat Professional converts a document from PDF into HTML or XML, the result is very nice. Both the HTML and XML formats are clean, relatively small, with a tagging structure and useful clues for processing. The results are consistent, satisfying the generality property. Therefore, our finalist candidates for input into logical structure extraction are HTML and XML formats as generated from PDF. To further narrow our choice of transformation, we analyzed the following sample parts of the targeted documents using the two finalist candidates. These cover an array of possible structures that appear repeatedly in OMG software specifications: (a) Sample paragraphs, (b) Sample figures, (c) Complex tables containing figures and hyperlinks in their cells, and (d) Nested lists with complicated hierarchy structures.

Our many assessments revealed that the XML format is more machine-readable and simple for analysis. Moreover, in an XML file, each tag is in a line, so we can analyze and parse the document line by line, which is easier compared to the HTML format in which we have to explore the document character by character. In the next section, our experimental outcomes related to the logical structure extraction are presented.

### 4. LOGICAL STRUCTURE EXTRACTION

After following the step described in previous section, we have the initial XML document. However, aspects of the document structure (such as headings) still need to be extracted to form a meaningful tagging structure in order to facilitate the further processing. The main motivation for further processing is that we found the documents tend to have consistent patterns of sentence structure and terminology in their document headings, various document body sections, and the index [Nojoumian 2007]. Our first assumption was that document headings (i.e., those that appear in the table of contents) carry the most important concepts with respect to a targeted document. This assumption seems particularly reasonable when we have a large document with numerous headings. That is why people usually explore the table of contents when they start working with a new document.

In the following section, we first discuss two implementation approaches to finalize our extraction of the logical structure. We then evaluate our methods and express the reasons for failure in the first technique. Finally, we present our successful practice for the logical structure extraction.

## 4.1 First Refinement Approach

In this approach, we applied a simple parser to scan for matching major tags, such as <Part>, <Sect> and <Div>, which Adobe Acrobat Professional used to open and close each part, chapter, section, etc of a document. Consider the following sample structure of a document (left-hand side). Using a straightforward stack-based parsing approach, we converted this into (right-hand side):

```
<Sect name="Generalization">                <Generalization>
    <Sect name="Class-Ref">                    <Class-Ref>
        <Sect name="Name">…</Sect>                <Name> … </Name>
        <Sect name="Package-Ref">…</Sect>         <Package-Ref> … </Package-Ref>
    </Sect>                                    </Class-Ref>
</Sect>                                    </Generalization>
```

Unfortunately, after running the program on different chapters and the whole document as well, it failed. We found out that there is a considerable amount of incorrect tagging. The tool opened each part, chapter, section, etc by <Sect> in a proper place in the document but it closed all of these tags by </Sect> in wrong places. The problem became more serious when we processed the whole document at once because of the accumulated mis-tagging. A sample of incorrect tagging is presented here:

```
<Sect number=" 7.3">
    <Sect number="7.3.1"> … </Sect>
    <Sect number="7.3.2"> … </Sect>
→ Correct place for closing <Sect number="7.3">
<Sect number="7.4">
</Sect>
</Sect>→ Wrong place
```

As a result, we could not extract the logical structure of the document in a meaningful format by this simple approach and decided to develop a new program that was more powerful and capable of detecting tagging errors. In the next section, our successful approach and the corresponding results are demonstrated.

## 4.2 Second Implementation Approach

In the second approach, we developed a more powerful parser that focused on a keyword, *LinkTarget*, which corresponds to the bookmark elements created in the previous transformation. This keyword is attached to each heading in the bookmark such as headers of parts, chapters, sections, and so on. Therefore, as a first step, we extracted all lines containing the mentioned keyword and put them in a queue, named *LinkTargetQueue*. We also defined various types of headings in the entire set of OMG specifications with respect to its logical structure. This classification is shown in Table 1.

**Table 1. Different kinds of headings**

| T | Sample Headings | Type |
|---|---|---|
| 1 | Part I - Structure | Part |
| 2 | 7 Classes | Chapter |
| 3 | 7.3 Class Descriptions | Section |
| 4 | 7.3.1 Abstraction | Subsection |
| 5 | Generalization, Notation, etc | Keyword |
| 6 | Annex | End part |
| 7 | Index | Last Part |

Then, we applied the ***Procedure LogicalStructureExtractor(LinkTargetQueue)*** that takes a queue as its input. Each node of this queue is a line of the input XML file which has *LinkTarget* string as a keyword, e.g., <P id="LinkTarget_111914">7 Classes </P>. This algorithm extracts headings (e.g., 7 Classes) and then defines their types by pattern matching according to Table 1 (e.g., $T_{Chapter} = 2$). Subsequently, it applies a stack-based approach for opening and closing corresponding tags at suitable places in the XML file. By applying this logical analyzer, we extracted all headings from various OMG software specifications and created new XML files for these documents with meaningful tags.

Fig. 1 shows a sample bookmark and its corresponding logical structure in XML format regarding one of the OMG specifications, i.e., UML Superstructure Specification (Unified Modeling Language). It consists of 4 major parts, 18 chapters, and numerous concepts such as generalizations, description, etc. We extracted 71 different types of tags in three categories (Structures, Blocks, and Keywords). Some of them, with their number of occurrence, are presented in Table 2. The general structure of these documents consists of parts, chapters, sections, subsections, and keyword-headed sub-subsections.

**Table 2. Sample XML tags in the UML superstructure specification**

| Structures | # | Blocks | # | Keywords | # |
|---|---|---|---|---|---|
| <Part> | 4 | <P>: Paragraph | 8228 | <Associations> | 177 |
| <Chapter> | 18 | <Figure>: Figure | 738 | <Attributes> | 171 |
| <Section> | 74 | <Table>: Table | 105 | <Constraints> | 172 |
| <Subsection> | 314 | <TH>:Table Header | 283 | <Description> | 202 |
| | | <TR>: Table Row | 547 | <Generalization> | 296 |
| | | <TD>: Table Data | 1721 | <Notation> | 169 |
| | | <L>: Lists | 245 | <Semantics> | 179 |
| | | <LI>: List Item | 765 | etc. | |

**Procedure** LogicalStructureExtractor(LinkTargetQueue)
    F **// a new XML file**
    L **// a line: e.g.: <P id="LinkTarget_111914">7 Classes </P>**
    H **// Heading: e.g.: 7 Classes**
    T **// Type: e.g.: for the Chapters, $T_{Chapter} = 2$**
    $T_{\text{Last member of the HeadingStack}} = 0$
    HeadingStack = empty
    **While** (LinkTargetQueue != empty) **do**
        **Get** "L" from the LinkTargetQueue
        **Extract** the heading "H" from the "L"
        **Define** heading's type: "T"
        **While** ($T =< T_{\text{Last member of the HeadingStack}}$) **do**
            **Pop** "H" and "T" from the HeadingStack
            **Close** the suitable tag w.r.t the popped "T"
            **If** (HeadingStack == empty)
                Break this while loop
            **End if**
        **End while**
        **Push** the new "H" and "T" in the HeadingStack
        **Open** new tags w.r.t the pushed "H" & "T"
    **End while**
    **While** (HeadingStack != empty) **do**
        **Pop** "H" and "T" from the HeadingStack
        **Close** the suitable tag w.r.t the popped "T"
    **End while**
    **Return** "F"
**End procedure**

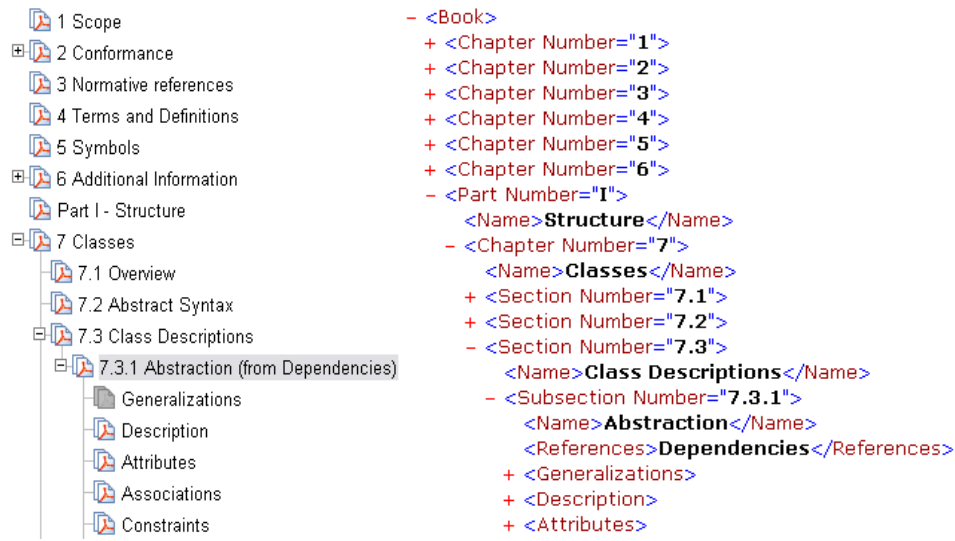

**Fig. 1. UML specification's bookmarks and its logical structure extracted in XML format**

# 5. TEXT EXTRACTION

Hypertext presentation has been a popular method for various computer applications dealing with large amounts of loosely structured information such as on-line documentation or computer-aided learning [Nielsen 1990b]. In this section, we take our results from the last section one step further in order to construct a general structure for our hypertext interfaces. We first evaluate whether each document is well formed and generates a valid schema. We then produce multiple HTML pages for each document while connecting them together. Finally, we demonstrate the construction of the document's key elements such as anchor links, figures, tables, and lists. The constructed multilayer hypertext versions consist of the following elements:

- A page for the table of contents.
- A separate page for each heading types, i.e., part, chapter, section, and subsection.
- Hyperlinks for accessing to the table of contents, next and previous pages.
- Some pages for extracted concepts, e.g., package and class hierarchy of the UML.
- Various cross references throughout the document.

To increase the usability of each document and highlight specific classes of information, we used different colors to present each XML element.

## 5.1 Checking Well-formedness and Validity

Every XML document must be well formed which means that it properly matches opening and closing tags and abides by logical rules of nesting. For well-formed checking and validating, we used a tool named *Stylus Studio® XML Enterprise Suite*, which is an XML integrated development environment.

In addition to checking for well-formedness, it is necessary for an XML document to be valid, i.e., whether a document uses tags in a consistent manner with its schema or not. A valid document has data that complies with a particular set of user-defined principles, or XML Schema, which illustrate correct data values and locations. Most of the XML tools support automatic schema generation in addition to the well-formedness checking. They also provide features for error detection during validation procedure, which makes it very easy to validate a schema. We first generated XML schemas and then validated the extracted documents.

## 5.2 Producing Multiple Outputs

To enhance the multilayer hypertext version's efficiency and facilitate document browsing, we produced multiple outputs using the *<xsl:result-document>* element, and generated a small hypertext page for each part, chapter, section, and subsection. The other alternative was to create a long HTML file (such as existing specifications on the web). Our motivation for generating small hypertext pages were as follows:

- **A better sense of location:** Users can have a better sense of location when navigating cross-references. In a large hypertext document one can use anchors (with the syntax <a name="xyz"> and <a href="#xyz">) to allow jumping from section to section. However, the result of jumping to a section in this manner places you into the middle of a document. Therefore, the user can find it confusing to determine exactly where they have arrived. On the other hand, if the destination of a jump is an entire hypertext page, the above problem goes away.
- **Less chance of getting lost**: Users are less likely to get lost by scrolling in small pages in comparison to a long page. In a long hypertext page, after following a link, a user may then move to some other parts of the document. But then the user may not know how to go back to where they came from unless they happen to remember the section number or title of the section they came from. If instead the document is organized as many small hypertext pages, it becomes simply a matter of hitting the *back* button in the browser.
- **A less-overwhelming sensation**: A smaller document should help users to manage larger amounts of information and understand the document more efficiently.
- **Faster loading**: Users are not always interested in downloading the whole document at once, especially when the document is fairly big.
- **Statistical analysis**: It may be useful to calculate the most frequent pages loaded and the time during which users stay in each page. This information could be used to improve the specification itself, and to determine what the most significant information is.

To prevent loss of the original order of a document, we created *Previous* and *Next* hyperlinks in each page in order to help the user to realize where he or she is, has been, and can go. It is important to note that there is a logical limit to how finely one wants to break down a large document into small hypertext pages (in the absurd extreme, one could separate each paragraph). What we have done is to limit the division to the subsection level.

In order to generate separate hypertext pages, we applied *Saxon*, which is an open source XSLT and XQuery processor developed by Michael Kay. Saxon versions exist for both .Net and Java. We used the Java version with the following command to transform the targeted XML documents by the XSLT code that we developed: "java -jar saxon8.jar -t filename.xml filename.xsl".

In our document, each part consists of a body as well as chapters, sections, and subsections inside of itself. Each chapter also consists of sections and subsections in addition to its body, and so forth. Therefore, to exclude chapters, sections, and subsections from an independent hypertext page, which just belongs to the body of a part, we had to create a global template for each of these entities in the XSLT code, as shown in Fig. 2. A global template is useful if an element occurs within various elements or in various locations of a document.

The other significant issue was the naming of these output files. This procedure had more importance when we wanted to link these hypertext files together and create the table of contents. Therefore, we used the following XPath function to name our outputs:

concat ('folder-name/', @Number, '.html')

This function concatenates three strings, creates a folder named "folder-name", and puts each hypertext file in this folder. The *@Number* refers to the attribute of <Part>, <Chapter>, <Section>, and <Subsection> elements. As a result, we named the hypertext outputs as follows: I.html, 7.html, 7.1.html, 7.2.html, 7.3.html, 7.3.1.html, 7.3.2.html, etc.

Since file names were created from the *@Number* attribute, we were able to facilitate access to each of these files. For instance, by a simple piece of XSLT code, as shown in Fig. 3, we generated the related hyperlinks in the table of contents.

```xml
<xsl:stylesheet xmlns:xsl="http://www.w3.org/1999/XSL/Transform" version="2.0">
  <xsl:output method="html" indent="yes" name="html" />
  <xsl:template match="/">
    <xsl:for-each select="//Part">
      <xsl:variable name="filename" select="concat('UML/',@Number,'.html')" />
      <xsl:value-of select="$filename" />
      <xsl:result-document href="{$filename}" format="html">
        <html>
          <body>
            <xsl:apply-templates />
            <!-- XSLT Codes -->
          </body>
        </html>
      </xsl:result-document>
    </xsl:for-each>
    <xsl:for-each select="//Chapter">
    <xsl:for-each select="//Section">
    <xsl:for-each select="//Subsection">
  </xsl:template>
  <xsl:template match="Chapter">
    <!-- Global Template -->
  </xsl:template>
  <xsl:template match="Section">
    <!-- Global Template -->
  </xsl:template>
  <xsl:template match="Subsection">
    <!-- Global Template -->
  </xsl:template>
</xsl:stylesheet>
```

**Fig. 2. Producing multiple outputs**

```xml
<a>
  <xsl:attribute name="href">
    <xsl:value-of select="concat(@Number,'.html')" />   → 7.3.html
  </xsl:attribute>
  <h4>
    <xsl:apply-templates select="@Number" />
    <xsl:text> </xsl:text>                              → 7.3 Class Descriptions
    <xsl:apply-templates select="Name" />
  </h4>
</a>
```

**Fig. 3. Generating hyperlinks in the ToC**

In the next section, we illustrate how to connect these files together by *Previous* and *Next* hyperlinks at the top of each page.

### 5.3 Connecting Hypertext Pages Sequentially

In the earlier section, we generated numerous hypertext pages for each OMG specification, for example, 418 pages for the UML Superstructure Specification. In a later section, we will be creating contextual hyperlinks and the table of contents that will allow direct jumping to various pages. However, we would still like to link all pages together by creating *Previous* and *Next* links in each page. This will allow the reader to proceed through the document in its original sequence, should they wish to do that. Therefore, we first extracted all elements' attribute, named *Number*, sequentially (1, 2, …, 7, 7.1, 7.2, 7.3, 7.3.1, etc) using a simple XSLT code. We then put them in a file, named *Num.txt*, and executed the algorithm ***Procedure Linker()*** in order to link hypertext pages together.

```
Procedure Linker()
    Num.txt // a text file consisting of all attributes
    A1, A2 // variables
    A1 = Read the first attributes from "Num.txt" file // (e.g. A1 = 1)
    A2 = Read the second attributes from "Num.txt" file // (e.g. A2 = 2)
    Call SetupLink (A1, A2) // (e.g. (1, 2))
    A1 = A2 // (e.g. A1 = 2)
    While (True) do
        A2 = Read an attribute from "Num.txt" // (e.g. A2 = 3, A2 = 4, A2 = 5)
        If (End of the "Nume.txt") Then
            Break this while loop
        End If
        Call SetupLink (A1, A2) // (e.g. (2, 3), (3, 4), (4, 5))
        A1 = A2 // (e.g. A1 = 3, A1 = 4, A1 = 5)
    End while
End procedure
Procedure SetupLink(X1, X2)
    folder-name Folder // a folder containing various hypertext files
    X1, X2 // arguments
    Extract the X1.html and X2.html from folder-name
    // (e.g. 7.3.1.html and 7.3.2.html)
    Set "Next" Hyperlink in X1.html based on the X2 variable
    // (e.g. in 7.3.1.html "Next" refers to 7.3.2.html)
    Set "Previous" Hyperlink X2.html based on the X1 variable
    // (e.g. in 7.3.2.html "Previous" refers to 7.3.1.html)
End procedure
```

In the next section, we demonstrate our presentation methods for different kinds of document major elements, and provide the related XSLT codes for the style sheet design.

## 5.4 Forming Major Document Elements

To construct the major document elements such as figures, tables, and lists, we developed various style sheets by XSLT programming and applied some tools such as *Altova StyleVision® Enterprise Edition* which is a visual style sheet designer for transforming XML and database content into HTML, PDF, and RTF output. In the next parts, a complete discussion with respect to the style sheet design for document elements is demonstrated with relevant XPath expressions and XSLT codes.

### 5.4.1 Figures

We first present the automatic extraction of a document's figures. In the transformation phase, when *Adobe Acrobat Professional* converted a document into an initial XML file, it also created a folder, named *images*, for the XML file. *Adobe* put all figures of the document in this folder, and named them as follows: folder-name_img_1.jpg to folder-name_img_n.jpg. The Fig. 4 shows the structure of the <Figure> element that has two children: (a) <ImageData> with its "src" attribute, and (b) <Caption>.

```
- <P>
  - <Figure>
      <ImageData src="images/UML_img_1.jpg" />
    - <Caption>
        <P>Figure 2.1 - Level 0 package diagram</P>
      </Caption>
    </Figure>
  </P>
```

**Fig. 4. Figure tag structure in the XML document**

For the relevant style sheet design, first we took out the targeted chapter (e.g., Chapter 2: Conformance) and extracted the <Figure> element. Then, we inserted a dynamic hyperlink inside of the *src* attribute by the following XSLT code and XPath expression:

<xsl: value-of select="string(.)"/>

This line of the code selects the value of the *string(),* which returns the string value of the argument. Here, it refers to the current node by *dot*. Therefore, it replaced the values of this attribute (i.e., images/folder-name_img_**1**.jpg … images/folder-name_img_**n**.jpg) into the hyperlinks, and imported all figures of each document into the right places inside of the document. We also imported the related captions to the end of each figure, Fig. 5.

```
- <xsl:template match="Figure">
  - <center>
    - <xsl:for-each select="ImageData">
      - <xsl:for-each select="@src">
        -                               →"images/UML_img_1.jpg"
          - <xsl:attribute name="src">
              <xsl:value-of select="string(.)" />
            </xsl:attribute>
          </img>
        </xsl:for-each>
      </xsl:for-each>
    </center>
  - <center>
    - <xsl:for-each select="Caption">
      - <xsl:for-each select="P">
          <xsl:apply-templates />
        </xsl:for-each>
      </xsl:for-each>
    </center>
  </xsl:template>
```

**Fig. 5. Dynamic importation of figures**

### 5.4.2 Tables

Now, we illustrate how to create a dynamic pattern for importing all tables with different sizes from the XML files corresponding to specifications. Fig. 6 shows the <Table> element structure. It has two children: (a) <Caption> element which consists of (a) plain text, and (b) <TR> element (Table Row) which has two different children: (b-1) <TH> element (Table Header), and (b-2) <TD> element (Table Data).

```
- <Table>
   - <Caption>
        <P>Table 2.1 Compliance statement</P>
     </Caption>
   - <TR>
        <TH>Compliance Summary</TH>
     </TR>
   - <TR>
        <TD>Level 1</TD>
        <TD>YES</TD>
        <TD>YES</TD>
        <TD>NO</TD>
     </TR>
```

**Fig. 6. Table tag structure in the XML document**

Dynamic table creation is supported by XSLT programming. In these tables, one of the dimensions is fixed and the other one is dynamic. For example, the number of columns is fixed but the number of rows is variable. To create a dynamic pattern for importing our tables, we first created the relevant caption, and then selected the <TR> element. Subsequently, we constructed all table cells.

To import table headers <TH> and table data <TD>, we applied the following XPath function: *position()*. This function returns the index position of the node that is currently being processed. As an example, consider the first <TR> element in the Fig. 6. If we apply *<TD> When: position() = 1 <TD>*, it returns *Level 1* string. We used each of the following expressions in a conditional branch through the first column to the last one, e.g., *position() = 1,…, position() = 6*. They imported relevant data into the related cells.

### 5.4.3 Lists

We now present the style sheet design for lists. Fig. 7 shows the <L> element structure for a simple list. It has two grandchildren: (a) <LI_Label>, and (b) <LI_Title>.

To present a simple list, we first extracted <LI_Label> and <LI_Title> elements by *<xsl:for-each select="LI_Label">* and *<xsl:for-each select="LI_Title">*, and then presented their contents. But for the nested lists, after extracting the second <L> element, we applied the following XPath expressions:

child :: * [position()=1] & child :: * [position()=2] → first second parts of the nested lists

```
- <P>
  - <L>
    - <LI>
        <LI_Label>●</LI_Label>
        <LI_Title>abstract syntax compliance.</LI_Title>
      </LI>
    - <LI>
        <LI_Label>●</LI_Label>
        <LI_Title>concrete syntax compliance .</LI_Title>
      </LI>
```

**Fig. 7. List tag structure in the XML document**

The *child::\** means select all children of the current node, and *child::\* [position()=1]* means select the child which is in the first place, and so forth, as shown in Fig. 8.

```
- <xsl:template match="L">
  - <xsl:for-each select="LI">
    - <ul>
      - <p style="text-align:justify;">
        - <xsl:for-each select="LI_Label">
          - <span style="color:navy; text-align:justify;">
              <xsl:apply-templates />
            </span>
          </xsl:for-each>
        - <xsl:for-each select="LI_Title">
          - <span style="color:navy; text-align:justify;">
              <xsl:apply-templates />
            </span>
          </xsl:for-each>
        </p>
      </ul>
    </xsl:for-each>
  - <xsl:for-each select="L">
    - <ul>
      - <p align="justify">
        - <span style="color:maroon;">
            <xsl:value-of select="child :: * [position()=1]" />
          </span>
        </p>
      - <ul>
        - <p align="justify">
          - <span style="color:olive;">
              <xsl:value-of select="child :: * [position()=2]" />
            </span>
          </p>
        </ul>
      </ul>
    </xsl:for-each>
  </xsl:template>
```

**Fig. 8. Importation of simple and nested lists**

## 6. CONCEPT EXTRACTION

In this section, we present a sample of concept extractions from targeted documents, specifically *OMG software modeling specifications* (these concepts include *class* and *package* hierarchies). We applied logical expressions in order to extract such concepts. Although this part has been designed for software modeling specifications, it can give us a general view of how to perform concept extraction from other documents.

As we mentioned in prior sections, there are numerous concepts in headings (this fact was one of our major reasons for the logical structure extraction of a document). As an example, Fig. 9 shows class descriptions with respect to the *Components* and *Composite Structures*. Using *from* as a keyword, it also presents the packages to which these classes belong. Since we tagged this information as chapter, section, and subsection headings, we extracted (using XPath expressions and XSLT code) the class and package hierarchies of modeling specifications in two separate pages.

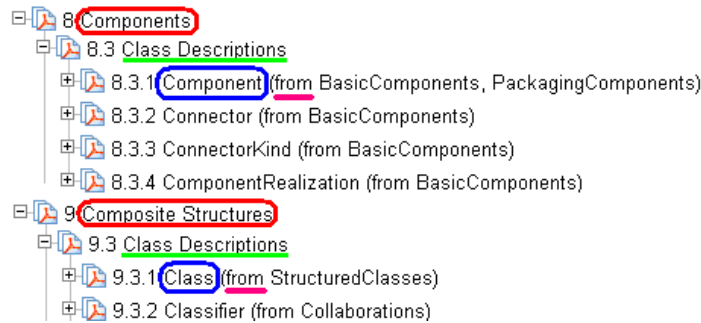

**Fig. 9. Headings containing UML concepts**

### 6.1 Modeling Class Hierarchy Extraction

In this part, we explain how to extract the class hierarchy from XML documents corresponding to modeling specifications. The main clue that we used in our extraction code was the *Class Descriptions*, which is a keyword string for the class hierarchy detection. For this reason, we applied the following XPath expression inside of the <Section> element (Fig. 10, arrow-I) to take out all classes:

child::*[position()=1 ]/starts-with(.,'Class Descriptions')

This expression means: select the first child of the <Section> element whose content starts with *Class Description*. By this logical expression, we only selected sections that present some descriptions about classes. Subsequently, we applied the following expression in order to define the title of a class set:

preceding-sibling :: * [last()]

This expression means: select the preceding sibling of the <Section> element in the last place (Fig. 10, arrow-III). As you can see in Fig. 10, <Section 9.3> has three preceding-siblings: <Section 9.2>, <Section 9.1>, and <Name> which is the last one. Finally, we moved to the <Subsection> element (Fig. 10 <Subsection 9.3.1>) and extracted contents of the <Name> element (e.g., *Class*) and the <Reference> element (e.g., *StructuredClasses*). We also linked this class to its relevant hypertext page by the <Subsection> element's attribute (i.e., @Number+html, for instance, 9.3.1.html).

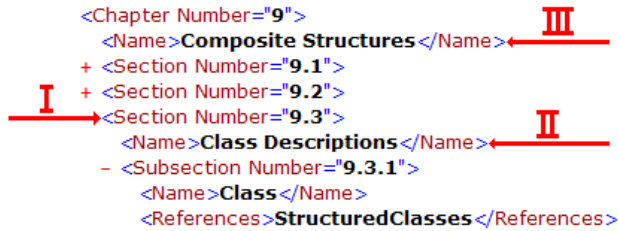

**Fig. 10. Part of tagging structures in the XML document**

As an example, part of the XSLT code with respect to the extraction of the UML class hierarchy is presented in Fig. 11.

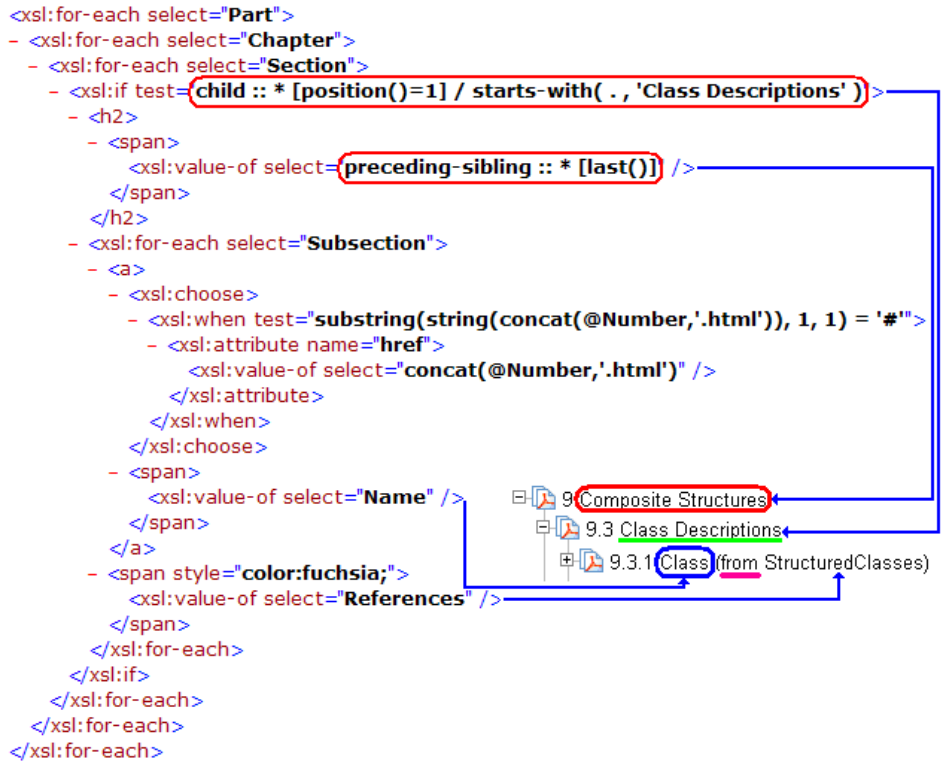

**Fig. 11. UML class hierarchy extraction**

## 6.2 Modeling Package Hierarchy Extraction

To extract the UML packages, we used the <Reference> element inside the <Subsection> element. The <Reference> element was created by *from* as a keyword string during the logical structure extraction. For instance, we applied the following expression inside of the <Subsection> element in order to extract all classes belonging to the *Actions* package:

contain(Reference,'Actions') = true() and
contain(Reference,'CompleteActions') = false() and …
contain(Reference,'StructuredActions') = false()

The *contain(string-1 , string-2)* function, returns true if string-1 contains string-2, otherwise, it returns false. Therefore, the above XPath expressions mean select subsections whose <Reference> element contains *Actions* but are not *CompleteActions* or *StructuredActions*, etc. As shown, we excluded other packages whose names overlapped with *Actions* package. Finally, we extracted the <Name> element, which carried the class names of the *Actions* package, and then linked each of these classes to its relevant hypertext page. Part of the XSLT code for the UML package hierarchy extraction is presented in Fig. 12.

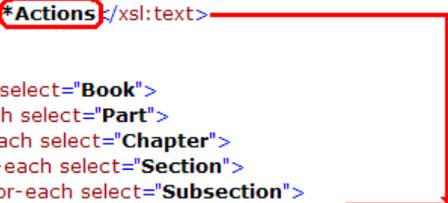

**Fig. 12. UML package hierarchy extraction**

We developed a simple script that could execute the above XSLT code repeatedly (plugging in each of the package names where *Actions* appears).

## 7. CROSS REFERENCING

To facilitate document browsing for end users, we created hyperlinks for major document keywords (for example, class names as well as package names) throughout the generated user interfaces. As we mentioned previously, since these keywords were among document headings, each of them had an independent hypertext page or anchor link in the final user interfaces. These hyperlinks help users to jump from one page to another page in order to gather more information as required.

We developed the related XSLT code to produce required strings for keywords used in the cross referencing algorithm, Fig. 13.

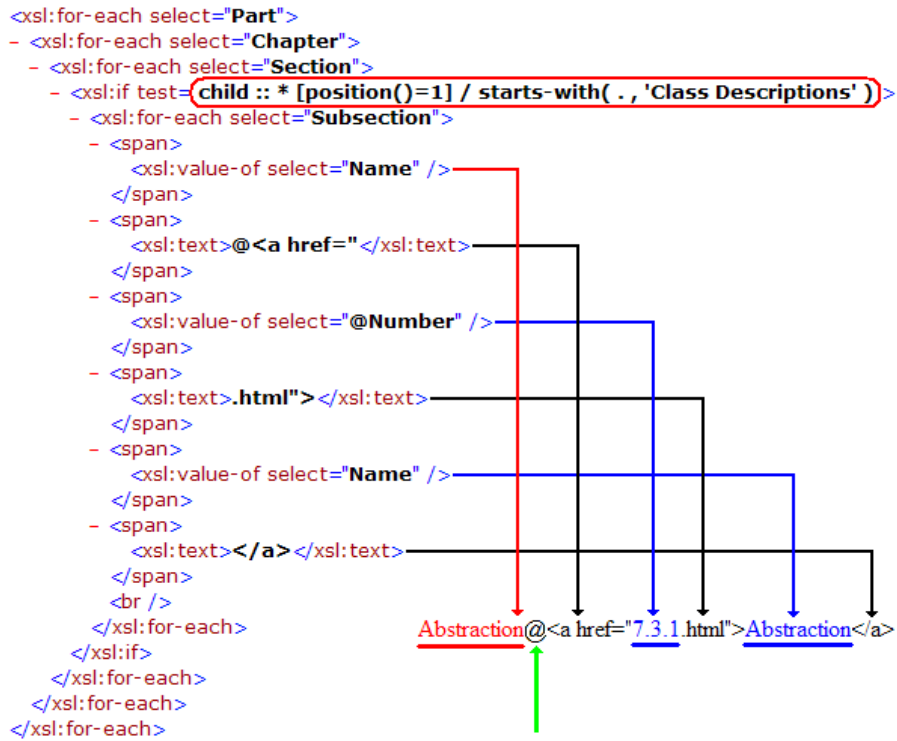

**Fig. 13. Producing related strings for cross-referencing**

This code selects sections that consist of class descriptions, and then generates a string which is made from the following six substrings, for every class:

Name+@<a href="+@Number+.html">+Name+</a>

For instance, *Abstraction* is a class name; therefore, its generated string is as follows:

Abstraction@<a href="7.3.1.html">Abstraction</a>

We applied a similar approach to generate related strings for package names, for example, the following string is generated for the *Actions* as a package name:

Actions@<a href="UMLPackage.html#Actions">Actions</a>

As you can see, we isolated keywords from their corresponding hyperlinks by @ character. We also listed all of these strings in a text file, named *UniqueKeywords.txt*, and then executed the ***Procedure CrossRef()*** for cross referencing.

To generalize this cross-referencing approach for other keywords and documents, we simply extracted all headers (since each had an independent hypertext page or anchor link) with their corresponding hyperlinks in order to put them in the *UniqueKeywords.txt* file, and then executed the *CrossRef* procedure.

```
Procedure CrossRef()
    folder-name // a folder consisting of various hypertext files
    F // a hypertext file belonging to a document
    UniqueKeywords.txt // a file consisting of the mentioned strings
    L // e.g.: Abstraction@<a href="7.3.1.html">Abstraction</a>
    S1, S2 // string variables
    While (True) do
        F = Extract a new hypertext page from folder-name
        If (all hypertext pages are extracted) Then
            Break this while loop
        Else
            While (end of the "UniqueKeywords.txt" file) do
                Get a new "L" from the text file // a new line
                Split "L" into two strings from "@" character
                S1 = first part of the "L" // Abstraction
                S2 = second part of the "L" // corresponding links
                If (find S1 in F in one place or many places) Then
                    Replace All (S1, S2) // replace all S1 strings with S2
                End If
            End while
        End If-Else
    End while
End procedure
```

## 8. EVALUATION, USABILITY, AND ARCHITECTURE

In this section, we demonstrate reengineering of various OMG software specifications, and address usability of generated multilayer hypertext versions by comparing them to the original PDF documents. We also illustrate the architecture of a document-engineering framework with the reengineering capability of PDF-based documents.

### 8.1 Reengineering of Various OMG Specifications

For further evaluation, we selected wide variety of other software specifications from Object Management Group (OMG) webpage with diverse number of pages and headings. The sample result of this assessment on ten documents is demonstrated in Table 3.

**Table 3. Sample reengineering of OMG specifications**

| Original OMG Specifications | Number of PDF Pages | Number of Headings | Headings Used in Cross-Ref | Number of Tokens in Doc Body | Number of Tokens in Headings | Data Analysis Results | Number of Hypertext Pages |
|---|---|---|---|---|---|---|---|
| CORBA | 1152 | 787 | 662 | 13179 | 702 | 15.1% | 788 |
| UML Sup. | 771 | 418 | 202 | 10204 | 378 | 12.2% | 421 |
| CWM | 576 | 550 | 471 | 6434 | 463 | 13.2% | 551 |
| MOF | 292 | 61 | 52 | 6065 | 92 | 8.0% | 62 |
| UML Inf. | 218 | 200 | 122 | 4329 | 176 | 9.3% | 201 |
| DAIS | 188 | 135 | 102 | 3051 | 151 | 12.6% | 136 |
| XTCE | 90 | 18 | 18 | 3075 | 26 | 2.6% | 19 |
| UMS | 78 | 69 | 59 | 1937 | 94 | 22.7% | 70 |
| HUTN | 74 | 88 | 83 | 2264 | 144 | 9.8% | 89 |
| WSDL | 38 | 17 | 17 | 1106 | 36 | 16.3% | 18 |

In this evaluation, for each of these documents, we created a separate hypertext page for its headings in addition to a page for its table of contents. To increase the usability of the outcomes, we did cross referencing all over hypertext pages by (a) detecting headings in each of these pages, and (b) connecting them to their corresponding entries. For instance, if the *AssociationClass* is among headings, it certainly has an independent hypertext page as well as hyperlinks in all the other pages where it appears. To avoid ambiguity, we filtered some phrases with common substrings (e.g., *Association* and *AssociationClass*), and eliminated phrases with many independent pages.

Furthermore, for each of these specifications, we sorted document and heading tokens based on their frequency in two separate lists. We then defined positions of heading tokens among document tokens, i.e., [$P_1$… $P_N$]. Finally, we determined how important the headings are. The data analysis column in Table 3 shows the headings are among the most frequent words, e.g., 15.1% shows headings are among top 15.1% frequent words in the entire document.

Fig. 14 also demonstrates the same concept with two different evaluations. In the lower diagram, we evaluated the headings whose number of occurrences were bigger that 2; but in the higher diagram we assessed the entire headings. As we mentioned earlier, this conclusion was our major motivation for:

- Extracting the logical structure based on the headings.
- Generating a separate hypertext page for each heading.
- Detecting major concepts among the document headings.
- Cross-referencing by detection of the document headings.

$$\mu: \text{Mean of } [P_1…P_N]$$
$$\partial: \text{Total number of document tokens}$$
$$\text{Percentage} = (\mu * 100) / \partial$$

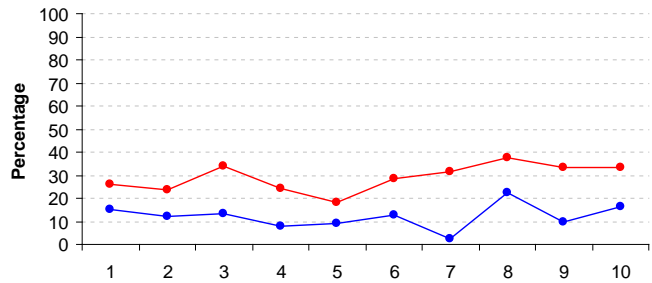

**Fig. 14. Headings are among the most frequent words**

All experiments confirmed that our approach is applicable to all kinds of documents. We just spent few seconds on some of these documents after the transformation phase to deal with rare mis-tagging problems. For example, forbidden notations among XML tags such as "**>**" (greater than) and "**<**" (less than) in some mathematical equations. Although this issue can be resolved automatically in our future design, the rest of our engineering procedures and software modules are totally automatic.

## 8.2 Usability of Multilayer Hypertext Interfaces

*Heuristic evaluation* is a systematic assessment of a user interface design in which a set of evaluators inspects the interface to judge its conformity with well-known usability principles [Nielsen and Molich 1990]. Nielsen [1989] compares 92 standard measurements of various usability issues related to hypertext in order to define those criteria that have the largest effects. Botafogo et al. [1992] also develop two types of metrics for hypertexts: global and node. The former refers to metrics concerning with the hypertext as a whole, and the latter focuses on the structural properties of individual nodes.

Although heuristic evaluation is not guaranteed to detect every single usability problem in an interface, this technique is a very efficient usability engineering method [Jeffries et al. 1991]. We first applied the same approach with the help of experts in our research lab. We then run a simple usability study among a group of software engineering students by designing multiple-choice questionnaires with an extra space for comments.

Our goal was to let them explore our user interfaces without any time limit such that they can also provide constructive feedback. For instance, they suggested that we add a Frame-like interface with a tree control on the left which shows the overall structure of a document, or create features that allow a user to add values to a document such as annotations, cross references, and links to related documentations.

In both methods, our intention was to compare the generated multilayer hypertext versions with the original PDF format as well as the HTML format of the specifications, which can be provided directly by *Adobe Acrobat*. This tool made a long hypertext page for each of those specifications along with anchors for headings at the top of each output.

Conklin [1987] summarizes operational benefits of hypertexts as follows: ease of tracing references, ease of creating new references, information structuring, global views in the ToC, customized documents, modularity, task stacking, and collaboration. Beside these advantages, we detected the following benefits through our usability studies, which did not exist in the original PDF formats, or *Adobe*-Generated HTML formats:

- **Navigating:** To be able to define previous, current, and next locations, and to go forward and backward by sequential browsing of headings. Indeed, the tendency to lose the sense of location and direction in a document is one of the major disadvantages of nonlinear hypertexts [Conklin 1987]. Moreover, a framework where users explore large amounts of information should have backtracking features in order to help the user to return directly to prior locations [Nielsen 1994].
- **Scrolling:** It would be confusing to scroll a long hypertext page containing hundreds of topics, headings, and cross-references rather than a set of small hypertext pages. Moreover, page boundary in the PDF version makes it difficult to follow up related materials spread over various pages (e.g., a big table or program code). Nielsen and Lyngbaak [1989] showed that 56 percent of the readers of a document presented in the hypertext format agreed with the statement "I was often confused about where I was".
- **Processing:** Accessing the structured content of PDF documents is a complex task, requiring reverse engineering and pre-processing techniques [Bloechle et al. 2006]. The generated hypertext interfaces simplify the content processing of the documents in the case of information retrieval, data mining, and knowledge acquisition.
- **Learning:** Humans can better handle a small amount of information presented in a single hypertext page, related to a unique topic. Based on the *minimize user memory load* principle, user interfaces should be simplified as much as possible, because every extra information or feature on a screen is one more thing for the user to learn [Nielsen 1994].
- **Monitoring:** To define a set of hypertext pages which have been downloaded several times and are probably more interesting for end users; they also get high ranking in popular search engines such as Google, Yahoo, etc.
- **Downloading:** Technical documents are not like novels. In other words, we do not need to provide the whole document at once. The better idea is to provide the table of contents as a menu for users, and let them to select whatever they require. In a large-scale assessment, this issue decreases network traffic a considerable amount.
- **Referencing:** Users should be able to jump to the desired location in a large information space. Therefore, a hypertext-like approach [Nielsen 1990a] with cross-referencing among various concepts, definitions, or even different documents would be a practical solution. For example, connecting UML Superstructure Specification to the UML Infrastructure Specification wherever it is necessary.
- **Coloring:** To be able to use different colors to present various classes of information and highlight some significant parts of the document automatically. It is a fact that some colors and color combinations are more visible than others [Durrett 1987].
- **Keeping track:** Users should be able to keep track of their interaction history and use this information subsequently [Greenberg 1993]. As a solution, colors of visited hyperlinks in the table of contents or other cross-references will be changed.

## 8.3 Architecture of the proposed framework

As we went further, by reengineering more specifications and technical documents, we modified our software components and ended up with the architecture of a document-engineering framework. It takes a PDF document and then generates corresponding meaningful XML format and multilayer hypertext version of the document. A company such as *Adobe* could use our approach in order to generate more useful versions of a document for both processing and browsing. This architecture is demonstrated in Fig. 15.

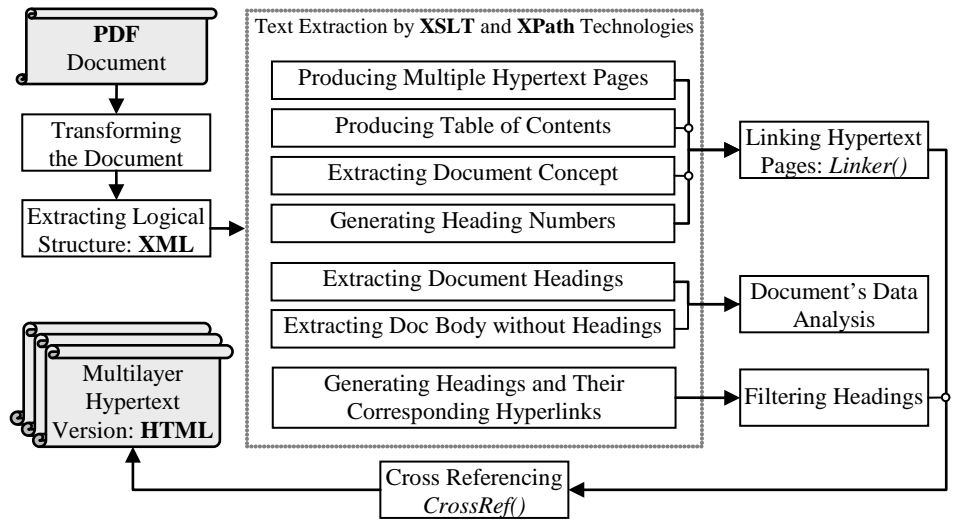

**Fig. 15. Architecture of the implemented document-engineering framework**

## 9. CONCLUSION AND FUTURE WORK

In this article, we described an approach for taking raw PDF versions of complex documents (e.g., specifications) and converting them into multilayer hypertext interfaces. For each document, we first generated a clean XML document with meaningful tags, and then constructed from this a series of hypertext pages constituting the final system.

The key contributions of the research are: (a) to illustrate methods for reengineering PDF-based technical documents in a general way, and (b) to demonstrate how to make a more usable hypertext version of documents so that end users to have a better experience with them. Our major goals were to make a complex document more usable by allowing navigation of both its structure and also of semantics described by the document.

We applied the latest W3C technologies such as XSLT and XPath expressions, and learned that, although by using these technologies we can parse every XML document, it would be more usable if the created XML documents have strong logical relationships among their elements and attributes similar to the XML documents we produced. As the final point, we propose research in the following directions as our future work:

- Extract the initial XML document from other formats such as DOC, RTF, HTML, etc. This can extend our framework for other kinds of formats and documents.
- Automate the concept extractions or at least create some features for the detection of the logical relationships among headings (as presented in Fig. 9).
- Improve the current solution and discover new users' demands. Only by such an investigation we can have a deep understanding of users' difficulties.

Access to all implementations and the reengineered OMG specifications are available.